\documentclass[twocolumn,showpacs,aps,prl,superscriptaddress,floatfix]{revtex4}

\usepackage{graphicx}
\usepackage{dcolumn}
\usepackage{amsmath}
\usepackage{epsfig}
 
\input pubboard/babarsym.tex

\newcommand{\BABARPubYear}     {05}
\newcommand{\BABARPubNumber}   {048}
\newcommand{\SLACPubNumber}    {11600}

\begin{document}
\noindent
\babar-PUB-\BABARPubYear/\BABARPubNumber\\
SLAC-PUB-\SLACPubNumber
\vskip 0.4cm

\title{
\large \bf  Search for the Rare Decays $B^0\to D_s^{(*)+} a_{0(2)}^-$}

\date{\today}% It is always \today, today, but you may specify any date with \date.

%% author list as of 02-Sep-2005 (633 authors)
%
\author{B.~Aubert}
\author{R.~Barate}
\author{D.~Boutigny}
\author{F.~Couderc}
\author{Y.~Karyotakis}
\author{J.~P.~Lees}
\author{V.~Poireau}
\author{V.~Tisserand}
\author{A.~Zghiche}
\affiliation{Laboratoire de Physique des Particules, F-74941 Annecy-le-Vieux, France }
\author{E.~Grauges}
\affiliation{IFAE, Universitat Autonoma de Barcelona, E-08193 Bellaterra, Barcelona, Spain }
% ...
\collaboration{The \babar\ Collaboration}
\noaffiliation

\begin{abstract}
We have searched for the decays
$B^0\to D_s^{+}a_0^-$,  $B^0\to D_s^{*+}a_0^-$, $B^0\to D_s^{+}a_2^-$ and  $B^0\to D_s^{*+}a_2^-$
in a sample of about 230 million
$\FourS\!\to\! B\Bbar$ decays collected with the \babar\ detector at the
\pep2\ asymmetric-energy \BF\ at SLAC. We find no evidence for these
decays and set upper limits at 90\% C.L. on the branching fractions:
${\cal B}(B^0\to D_s^+ a_0^-) < 1.9\times 10^{-5}$,
${\cal B}(B^0\to D_s^{*+} a_0^-) < 3.6\times 10^{-5}$,
${\cal B}(B^0\to D_s^+ a_2^-) < 1.9 \times 10^{-4}$,
and 
${\cal B}(B^0\to D_s^{*+} a_2^-) < 2.0\times 10^{-4} $.\\
\end{abstract}

% PACS, the Physics and Astronomy Classification Scheme.

\maketitle

The time-dependent decay rates for neutral $B$ mesons into a $D$
meson and a light meson provide sensitivity to the Cabibbo-Kobayashi-Maskawa (CKM)~\cite{ckm} 
quark mixing matrix phases $\beta$
and $\gamma$~\cite{angles}. A \CP-violating term emerges through the interference
between $B^0 \Bbar^0$ mixing mediated and direct decay amplitudes. 
The time-dependent \CP-asymmetries in the decay modes $B^0\to  D^{(*)-}
\pi^+$~\cite{chargeconj} have been studied by \babar\ and
BELLE~\cite{2b+g-expts-1, 2b+g-expts-2}. In these modes, the \CP-asymmetries arise due to
a phase difference between two amplitudes of very different magnitudes:
one decay amplitude is suppressed by the product of two small CKM
elements $V_{ub}$ and $V_{cd}$, while the other is CKM
favored. Therefore, the decay rate is dominated by the CKM-favored part
of the amplitude, resulting in a very small \CP-violating asymmetry. 

\begin{figure}[h]
\begin{center}
\begin{minipage}[h]{4.2cm}
\epsfysize=2.05cm
\epsfbox{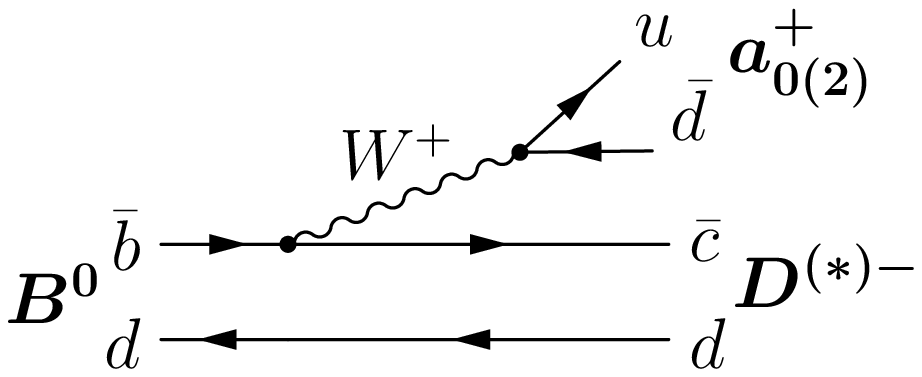}
\end{minipage}
\begin{minipage}[h]{4.2cm}
\epsfysize=2.05cm
\epsfbox{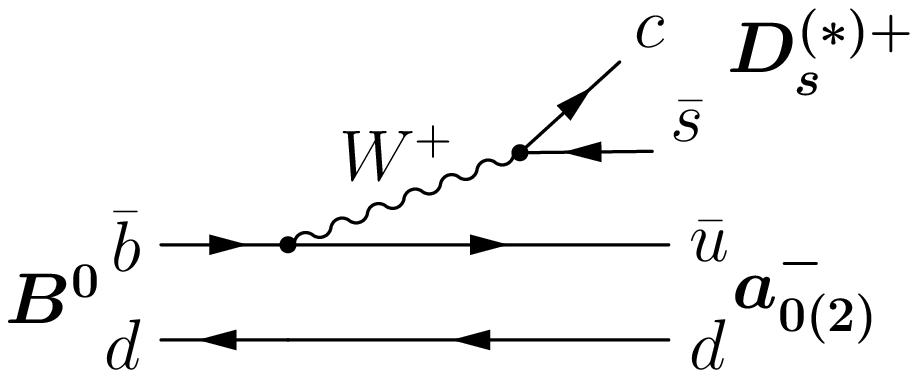}
\end{minipage}
\begin{minipage}[h]{4.2cm}
\epsfysize=2.05cm
\epsfbox{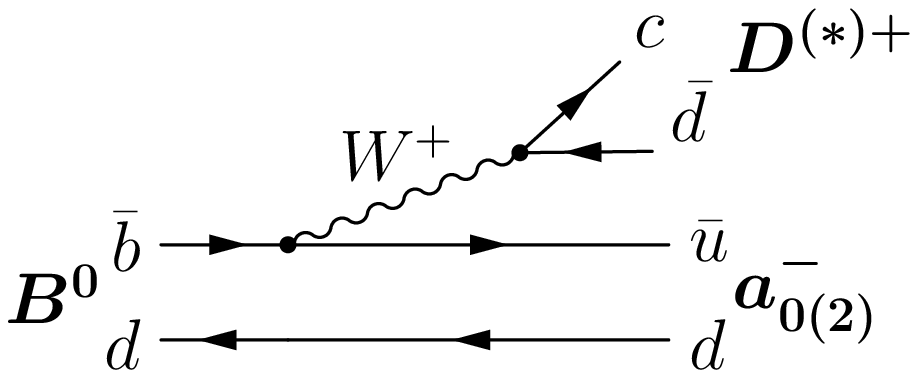}
\end{minipage}
\end{center}
\caption{Top diagrams: tree diagrams contributing to the decay amplitude of $B^0 \ra D^{(*)-} a^+_{0(2)}$ 
(including the $B^0 \Bbar^0$ mixing mediated part of the amplitude). Bottom diagram: 
tree diagram representing the decay amplitude of $B^0 \ra D^{(*)+}_s a^-_{0(2)}$.}
\label{fig:intro}
\end{figure}

Recently it was proposed to consider other types of light mesons in the
two-body final states~\cite{yet-another-way}. The idea is that decay
amplitudes with light scalar or tensor mesons, such as $a_0^+$ or
$a^+_2$, emitted from a weak current, are significantly suppressed
because of the small coupling constants $f_{a_{0(2)}}$. In the $SU(2)$
limit, $f_{a_0} = 0$ (since the coupling constant of a light scalar is
proportional to the mass difference between $u$ and $d$ quarks), and any
non-zero value of $f_{a_0}$ is of the order of isospin conservation breaking effects.  
Since the light tensor meson $a_2^+$ has spin 2, it cannot be emitted by
a $W$-boson (i.e.\ $f_{a_2} \equiv 0$), and thus could only appear in a
$V_{cb}$-mediated process via final state hadronic interactions and
rescattering.  
Therefore, the absolute values of the CKM-suppressed and favored parts
of the decay amplitude (see Figure~\ref{fig:intro}, top two diagrams) could become 
comparable, potentially resulting in a large \CP-asymmetry. 
No $B \ra a_{0(2)} X$ transitions have been observed yet.
A summary of the theoretical predictions for the values of $V_{ub}$ and $V_{cb}$-mediated
parts of the $B^0 \ra D^{(*)-} a_{0(2)}^+$ branching fractions
can be found in~\cite{dh}.

The $V_{ub}$-mediated amplitudes in~\cite{dh} were
computed in the factorization framework. In addition to model
uncertainties,  significant uncertainty in the theoretical calculations
is due to unknown $B \ra a_{0(2)} X$ transition form factors.
One way to verify the numerical assumptions and test the validity of
the factorization approach experimentally is to measure the branching
fractions for the $SU(3)$ conjugated 
decay modes $B^0 \ra D^{(*)+}_s
a_{0(2)}$. These decays are represented by a single tree diagram
(Figure~\ref{fig:intro}, bottom diagram) with external $W^+$ emission, without
contributions from additional tree or penguin diagrams. The
$V_{ub}$-mediated part of the $B^0 \ra D^{(*)+} a_{0(2)}^-$ decay
amplitude can be related to $B^0 \ra D^{(*)+}_s a_{0(2)}^-$ using
$\tan{(\theta_{\rm Cabibbo})} = |V_{cd}/V_{cs}|$ and the ratio
of the decay constants $f_{D_s^{(*)}}/f_{D^{(*)}}$.  

Branching fractions of $B^0 \ra D^{(*)+}_s a_2^-$ are predicted to be
in the range 1.3--1.8 (2.1--2.9) in units of $10^{-5}$~\cite{klo}. Branching fraction estimates for 
$B^0 \ra D^{(*)+}_s a_0^-$ of approximately $8 \times 10^{-5}$ are obtained using $SU(3)$ 
symmetry from the 
predictions made for $B^0 \ra D^{(*)+} a_0^-$ in~\cite{dh}.

In this paper we present the first search for the decays  $B^0\to
D_s^{+}a_0^-$, $B^0\to D_s^{*+}a_0^-$, $B^0\to D_s^{+}a_2^-$ and  $B^0\to D_s^{*+}a_2^-$. 
The analysis uses a sample of approximately $210~$\invfb, which corresponds to 
about 230 million \FourS decays into
$B \Bbar$ pairs collected in the years 1999--2004 with the \babar\
detector at the asymmetric-energy $B$-factory PEP-II~\cite{pep2}. The \babar\
detector is described elsewhere~\cite{babar} and only the components crucial 
to this analysis are summarized here. Charged particle tracking is
provided by a five-layer silicon vertex tracker (SVT) and a 40-layer
drift chamber (DCH). For charged-particle identification, ionization
energy loss ($dE/dx$) in the DCH and SVT, and Cherenkov radiation
detected in a ring-imaging device are used. Photons are identified and
measured using the electromagnetic calorimeter, which is comprised of 6580
thallium-doped CsI crystals. These systems are located inside a 1.5~T 
solenoidal superconducting magnet. We use GEANT4~\cite{geant} software to
simulate interactions of particles traversing the \babar\ detector,
taking into account the varying detector conditions and beam
backgrounds.

The selection criteria are optimized by maximizing the ratio of
expected signal events $S$ to the square-root of the sum of
signal and background events $B$. For the calculation
of $S$ we assume ${\cal B}(B^0\to D_s^{(*)+}a_2^-)$ to be the mean 
values of the predicted intervals from~\cite{klo} and
an estimate of ${\cal B}(B^0\to D_s^{(*)+}a_0^-)$ is obtained from
${\cal B}(B^0\to D^{(*)+}a_0^-)$  predicted in~\cite{dh} and assuming $SU(3)$ symmetry. 
The optimal selection
criteria as well as the shapes of the distributions of selection
variables are determined from simulated Monte Carlo (MC) events. We use
MC samples of our signal modes and, to simulate background, inclusive samples of 
$B^+B^-$ (800~fb$^{-1}$), 
$B^0 \Bbar^0$ (782~fb$^{-1}$),   
$c\bar{c}$ (263~fb$^{-1}$), and   
$q\bar{q},~q = u,d,s$ (279~fb$^{-1}$). In addition, we use large samples of simulated
events of rare background modes 
which have final states similar to the signal.

Candidates for \Ds mesons are reconstructed in the modes $\Ds\to
\phi\pi^+$, $\Kstarzb K^+$, and $\KS K^+$, with $\phi\to K^+K^-$,
$\Kstarzb\to K^-\pi^+$ and $\KS\to\pi^+\pi^-$. 
The \KS candidates
are reconstructed from two oppositely-charged tracks, with an invariant
mass close to the nominal \KS mass~\cite{PDG2004}, that come from a
common vertex displaced from the $e^+e^-$ interaction point.
All other tracks are required to originate less than 1.5~cm away from the
$e^+e^-$ interaction point in the transverse plane and less than 10~cm
along the beam axis.
Charged kaon candidates must satisfy kaon identification criteria
that are typically around 95\% efficient, depending on momentum and
polar angle, and have a misidentification rate at the 10$\%$ level.
The $\phi\to K^+K^-$, 
$\Kstarzb\to K^-\pi^+$ and $\KS\to\pi^+\pi^-$ candidates are required to have invariant
masses close to their nominal masses~\cite{PDG2004} 
(we require the absolute differences
between their measured masses and the nominal values~\cite{PDG2004} to be in the range
12--15~\mev,
35--60~\mev and 7--12~\mev, respectively,
depending on the $B^0$ and $D^+_s$ decay modes).
The polarizations of the \Kstarzb and $\phi$ mesons in the \Ds decays are
used to reject backgrounds through the use of the helicity
angle $\theta_H$, defined as the angle between the $K^-$ momentum vector and the
direction of flight of the \Ds in the 
\Kstarzb or $\phi$ rest frame. 
The \Kstarzb candidates are required to have $|\cos\theta_H|$ greater than 0.25--0.5
and $\phi$ candidates are
required  to have $|\cos\theta_H|$ greater than 0.3--0.5, depending on the $B^0$ decay mode. 
We also apply a vertex fit to the $D^+_s$ candidates that decay into $\phi \pi^+$
and $\Kstarzb K^+$, since all charged daughter tracks of $D^+_s$ are supposed to
come from a common vertex. The $\chi^2$ of the vertex fit is required to be less than
10--16 (which corresponds to a probability of better than 
$0.1\% - 1.9\%$ for the 3 track vertex fit), depending on the reconstructed mode.

The $D^{*+}_s$ candidates are reconstructed in the mode $D^{*+}_s \ra D^+_s \gamma$.
The photons are required to have an energy greater than 100~\mev.
The \Ds and $D^{*+}_s$ candidates are
required to have invariant masses less than about $\pm 2\sigma$
from their nominal values~\cite{PDG2004}. 
The invariant mass of the
$D^{*+}_s$ is calculated after the mass constraint on the daughter $D^+_s$
has been applied.
Subsequently, all $D^{*+}_s$ candidates
are subjected to a mass-constrained fit. 

We reconstruct $a_0^-$ and $a_2^-$ candidates in their decay to the
$\eta \pi^-$ final state. For reconstructed $\eta\to\gamma\gamma$
candidates we require the energy of each photon to be greater than
250~\mev for $a_0^+$ candidates, and greater than 300 -- 400~\mev for
$a_2^+$ candidates, depending on the $D^+_s$ mode. 
The $\eta$
mass is required to be within a $\pm 1\sigma$ or $\pm 2\sigma$ interval of the nominal value~\cite{PDG2004},
depending on the background conditions in a particular $B^0$, $D^+_s$ decay mode
(the $\eta$ mass resolution is measured to be around 15~\mevcc).
The $a_0^+$ and $a_2^+$ candidates are required to have a mass 
$m_{\eta \pi^+}$ in the range 0.9--1.1~\gevcc and 1.2--1.5~\gevcc,
respectively.
We also require that photons from $\eta$ and $D^{*+}_s$ are inconsistent 
with $\pi^0$ hypothesis when combined with any other photon in the event
(the $\pi^0$ veto window varies from $\pm 10$ to $\pm 15$~\mevcc). 
Finally, the $B^0$ meson candidates are formed using the reconstructed
combinations of $D^+_s a_0^-$, $D^+_s a_2^-$, $D^{*+}_s a_0^-$
and $D^{*+}_s a_2^-$.

The background from continuum $q\bar{q}$ production (where $q = u,d,s,c$) is
suppressed based on the event topology. We calculate the angle ($\theta_T$)
between the thrust axis of the $B$ meson candidate and the thrust axis
of all other particles in the event. In the center-of-mass frame (c.m.),
$B\Bbar$ pairs are produced approximately at rest and have a 
uniform $\cos\theta_T$ distribution. In contrast, $q\bar{q}$ pairs are
produced in the c.m.\ frame with high momentum, which results in a
$|\cos\theta_T|$ distribution peaking at 1. Depending on the background
level of each mode, $|\cos\theta_T|$  is required to be smaller than
0.70--0.75. We further suppress backgrounds using a Fisher discriminant
(${\cal F}$)~\cite{fisher} constructed from the scalar sum of the c.m.\ momenta of all
tracks and photons (excluding the $B$ candidate decay products) flowing
into 9 concentric cones centered on the thrust axis of the $B$
candidate. The more isotropic the event, the larger the value of ${\cal F}$. 
We require ${\cal F}$ to be larger than a threshold that retains 
$75\%$ to $86\%$ of the signal while rejecting $78\%$ to $65\%$ of
the background, depending on the background level.
In addition, the
ratio of the second and zeroth order Fox-Wolfram
moments~\cite{fox-wolfram} must be less than a threshold in the
range 0.25--0.40 depending on the decay mode.  

We extract the signal using the kinematical variables $\mes =
\sqrt{E_{\rm b}^{*2} - (\sum_i {\mathbf p}^*_i)^2}$ and $\Delta E =
\sum_i\sqrt{m_i^2+{\mathbf p}_i^{*2}} - E_{\rm b}^*$, where $E_{\rm b}^*$ is
the beam energy in the c.m.\ frame, ${\mathbf p}^*_i$ is the c.m.\
momentum of the daughter particle $i$ of the $B^0$ meson candidate, and
$m_i$ is the mass hypothesis for particle $i$. For signal events, \mes
peaks at the $B^0$ meson mass with a resolution of about 2.7~\mevcc and
$\Delta E$ peaks near zero with a resolution of 20~MeV, indicating that
the $B^0$ candidate has a total energy consistent with the beam energy in
the c.m.\ frame. The $B^0$ candidates are required to have $|\Delta E|<
40\ \mev$ and $\mes > 5.2~\gevcc$.

The fraction of multiple $B^0$ candidates per event is estimated using the
MC simulation and found to be around $2\%$ for $D^+_s a_{0(2)}^-$
and $5\%$ for $D^{*+}_s a_{0(2)}^-$ combinations. In each event with
more than one $B^0$ candidate that passed the selection requirements, we select
the one with the lowest $|\Delta E|$ value.

After all selection criteria are applied, we estimate the $B^0$
reconstruction efficiencies, excluding the intermediate branching fractions
(see Table~\ref{tab:eff}).

\begin{table}[!h]
\begin{center}
\caption{Reconstruction efficiencies for $B^0 \ra D^{(*)+}_s a_{0(2)}^-$ decays (excluding the intermediate
branching fractions).}
\label{tab:eff}
\vspace{\baselineskip}
\begin{tabular}{ c c c c c }
\hline\\[-7pt]
 Decay mode  & $D_s^+ \ra \phi \pi^+$~ & $D^+_s \ra \Kstarzb K^+$~ & 
$D^+_s \ra \KS K^+$ \\
\hline\\[-7pt]
$B^0 \ra D^+_s a_{0}^-$ & 4.7$\%$ & 2.9$\%$ & 2.5$\%$ \\
\\[-7pt]
$B^0 \ra D^+_s a_{2}^-$ & 1.9$\%$ & 1.1$\%$ & 1.1$\%$ \\
\\[-7pt]
$B^0 \ra D^{*+}_s a_{0}^-$ & 2.2$\%$ & 1.5$\%$ & 1.3$\%$ \\
\\[-7pt]
$B^0 \ra D^{*+}_s a_{2}^-$ & 0.9$\%$ & 0.7$\%$ & 0.5$\%$ \\[2pt]
\hline
\end{tabular}
\end{center}
\end{table}

Background events that pass these selection criteria are mostly
from $q\bar{q}$ continuum, and their \mes distribution is described by a
threshold function~\cite{Argus}: 
\begin{equation*}
f(m_{\rm ES}) \sim m_{\rm ES} \sqrt{1-x^2} {\rm exp}[-\xi (1-x^2)],
\label{eq:argus}
\end{equation*}
where $x = 2 m_{\rm ES}/\sqrt{s}$, $\sqrt{s}$ is the total energy
of the beams in their center of mass frame, and $\xi$ is the
fit parameter. 
A study using simulated events of $B^0$ and $B^+$
decay modes with final states similar to our signal mode, including
$D^{(*)+}_s \pi^-$ and $D^{(*)+}_s \rho^-$, shows that these modes do not
peak in \mes.

\begin{figure}[!h] %
  \begin{center}%
\epsfig{figure=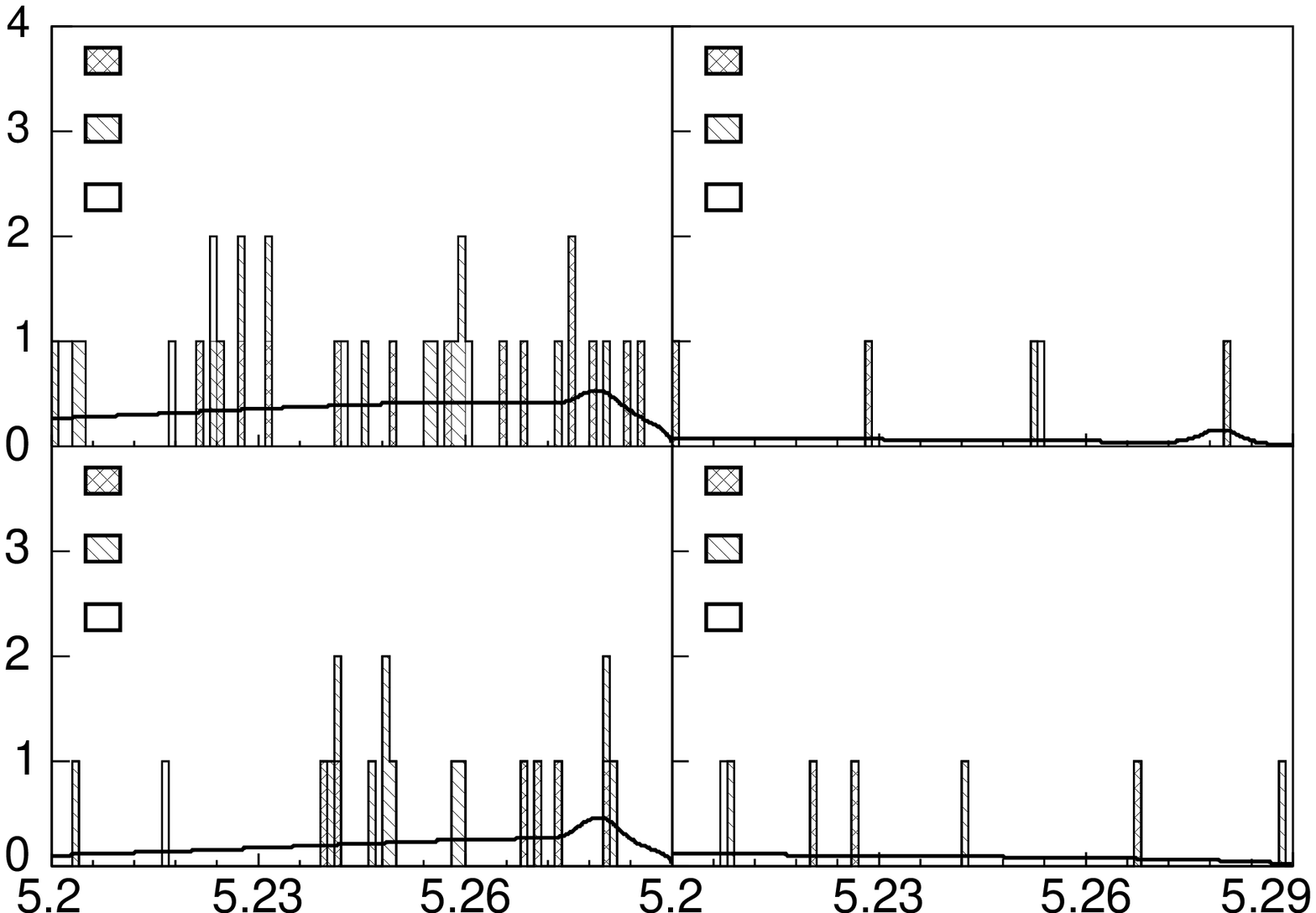,bbllx=0,bblly=0,bburx=555,bbury=380,width=1.00\linewidth}%
    \put(-168,140){\large \boldmath $D^+_s a_0^-$}
    \put(-218,153){\small $\phi \pi^+$}
    \put(-218,140){\small $\Kstarzb K^+$}
    \put(-218,128){\small $\KS K^+$}
    \put(-50,140){\large \boldmath $D^+_s a_2^-$}
    \put(-104,153){\small $\phi \pi^+$}
    \put(-104,140){\small $\Kstarzb K^+$}
    \put(-104,128){\small $\KS K^+$}
    \put(-168,60){\large \boldmath $D^{*+}_s a_0^-$}
    \put(-218,76){\small $\phi \pi^+$}
    \put(-218,63){\small $\Kstarzb K^+$}
    \put(-218,51){\small $\KS K^+$}
    \put(-50,60){\large \boldmath $D^{*+}_s a_2^-$}
    \put(-104,76){\small $\phi \pi^+$}
    \put(-104,63){\small $\Kstarzb K^+$}
    \put(-104,51){\small $\KS K^+$}
    \put(-75,-11){\small \boldmath \bf ${\rm m_{\rm ES}~~(GeV/c^2)}$}
    \put(-255,55){\rotatebox{90}{\small \boldmath \bf ${\rm Events/1~MeV/c^2}$}}
    \caption{
      Distributions of \mes\ for $B^0\to D_s^{(*)+} a_{0(2)}^-$ candidates
      overlaid with the projection of the maximum likelihood fit. Contributions from $D^+_s$ modes
      are shown with a different hatching style. The fit procedure and results are
      described in the text.}    
    \label{fig:mes-all}
  \end{center}
\end{figure}

Figure~\ref{fig:mes-all} shows the \mes
distributions for the reconstructed candidates $B^0 \to D_s^{+} a_0^-$, $B^0 \to D_s^{+} a_2^-$,
$B^0 \to D_s^{*+} a_0^-$ and  $B^0\to
D_s^{*+}a_2^-$. For each mode, we perform an
unbinned maximum-likelihood fit to the \mes distributions using the
candidates from all \Ds decay modes combined. We fit the \mes
distributions with the 
sum of the function $f(m_{ES})$ characterizing
the combinatorial background and a Gaussian function to describe the
signal. 
The total signal yield in each $B^0$ decay mode is calculated as
a sum over $D^+_s$ modes ($i=\phi \pip$, $\Kstarzb K^+$, $\KS K^+$): 
\begin{equation*}
n_{sig} = {\cal B} \cdot N_{B \bar B} \cdot \sum_i  {\cal B}_i \cdot \epsilon_i, 
\end{equation*}
where 
${\cal B}$ is the branching fraction of the $B^0$ decay mode, 
$N_{B \bar B}$ is the number of produced $B\bar B$ pairs,
${\cal B}_i$ is the product of the intermediate
branching ratios and $\epsilon_i$ is the reconstruction efficiency.
The mean and the width of the Gaussian function are fixed to
values obtained from simulated signal events for each decay mode. The threshold
shape parameter $\xi$, along with the branching ratio ${\cal B}$ 
are free parameters of the fit.
The likelihood function is given by:
\begin{equation*}
{\cal L} = \frac{e^{-N}}{N!} \prod_{i=1}^N (n_{sig} P_i^{sig} + (N-n_{sig}) P_i^{bkg}),
\label{eq:likelihood}
\end{equation*}
where 
$P^{sig}_i$ and $P^{bkg}_i$ are the probability density functions
for the corresponding hypotheses, $N$ is the total number of events in the fit
and $i$ is the index over all events in the fit.

Table~\ref{tab:fit} (second column) shows the signal event yields from the \mes fit.
Due to a lack of entries in the signal region for the $B^0 \to D_s^{*+} a_2^-$ mode,
the fit did not yield any central value for the number of signal
events in this mode. 
Accounting for the
estimated reconstruction efficiencies and daughter particles branching fractions,
we measure the branching fractions shown in the third column of Table~\ref{tab:fit}.

\begin{table}[!h]
\begin{center}
\caption{Signal yields, branching fractions
and upper limits on the branching fractions 
for $B^0 \ra D^{(*)+}_s a_{0(2)}^-$ decays.
Numbers in parentheses in the third and fourth columns indicate the branching
fractions and the upper limits
multiplied by the branching fractions of the decays $D^+_s \ra \phi \pi^+$ 
and $a_{0(2)}^+ \ra \eta \pi^+$.}
\label{tab:fit}
\vspace{\baselineskip}
\begin{tabular}{ l c c c c }
\hline\\[-7pt]
$B^0$ mode  & $n_{sig}$~ & ~${\cal B}~[10^{-5}(10^{-7})]$ & 
$U.L.\ ~[10^{-5}]$ \\
\hline\\[-7pt]
$D^+_s a_{0}^-$ & $0.9^{+2.2}_{-1.7}$ & $0.6^{+1.4}_{-1.1} \pm 0.1~~(2.6^{+6.6}_{-5.1} \pm 0.5)$ & $1.9~(0.09)$ \\
\\[-7pt]
$D^+_s a_{2}^-$ & $0.6^{+1.0}_{-0.6}$ & $6.4^{+10.4}_{-5.7} \pm 1.5~~(4.5^{+7.3}_{-4.0} \pm 0.8)$ & $19~(0.13)$ \\
\\[-7pt]
$D^{*+}_s a_{0}^-$ & $1.5^{+2.3}_{-1.8}$ & $1.4^{+2.1}_{-1.6} \pm 0.3~~(6.5^{+10.1}_{-7.8} \pm 1.2)$ & $3.6~(0.17)$ \\
\\[-7pt]
$D^{*+}_s a_{2}^-$ & $-$ & $- ~~(-)$ & $20~(0.13)$ \\[2pt]
\hline
\end{tabular}
\end{center}
\end{table}

The systematic errors include a 14\% relative uncertainty for
$D_s^+$ decay rates~\cite{dsphipi-babar}. Uncertainties in the \mes 
signal and background shapes result
in 11\% relative error in the measured branching fractions. 
The rest of the systematic error sources, which include uncertainties in
photon and $\eta$ reconstruction efficiencies, the $a_0^+$ and $a_2^+$ masses and
widths, track and \KS reconstruction, charged kaon identification, range 
between 3\% and 10\%. We assume the branching fraction for $a_0^+\to
\eta\pi^+$ to be 100\% and assign an asymmetric systematic error of
$-10\%$ to this assumption. The systematic error in the number of
produced $B\Bbar$ pairs is 1.1$\%$.
It was checked that the selection of the best candidate based
on $|\Delta E|$ does not introduce any significant bias
in the $m_{ES}$ fit.
The total relative systematic errors are estimated to be around $25\%$ for each mode.

\begin{figure}[!h] %
  \begin{center}%
\epsfig{figure=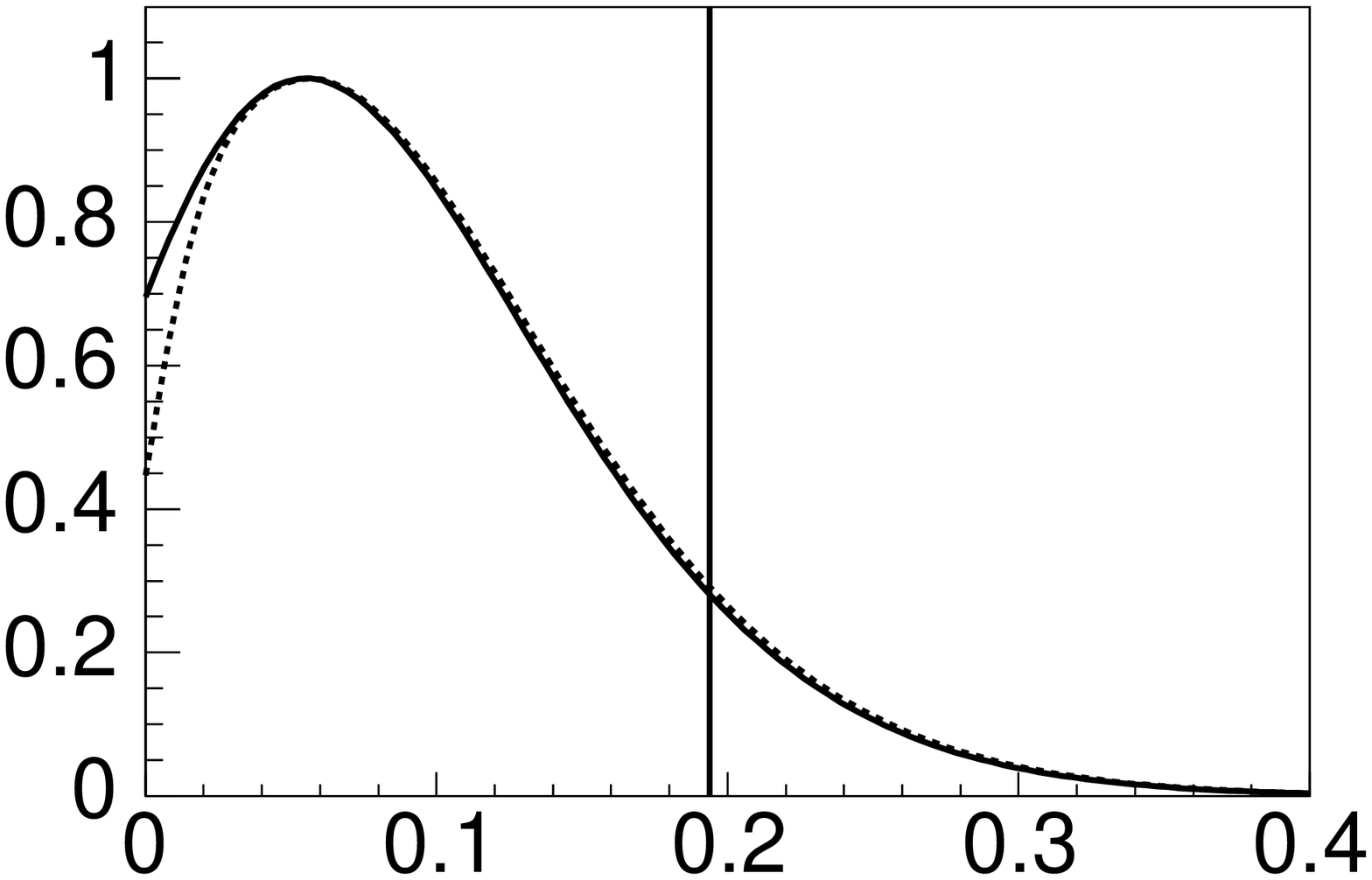,bbllx=0,bblly=0,bburx=555,bbury=380,width=0.51\linewidth}%
    \put(-55,60){\boldmath \large $D^+_s a_0^-$}
    \put(-40,-7){\boldmath \bf ${\cal B} \times 10^4$}
\epsfig{figure=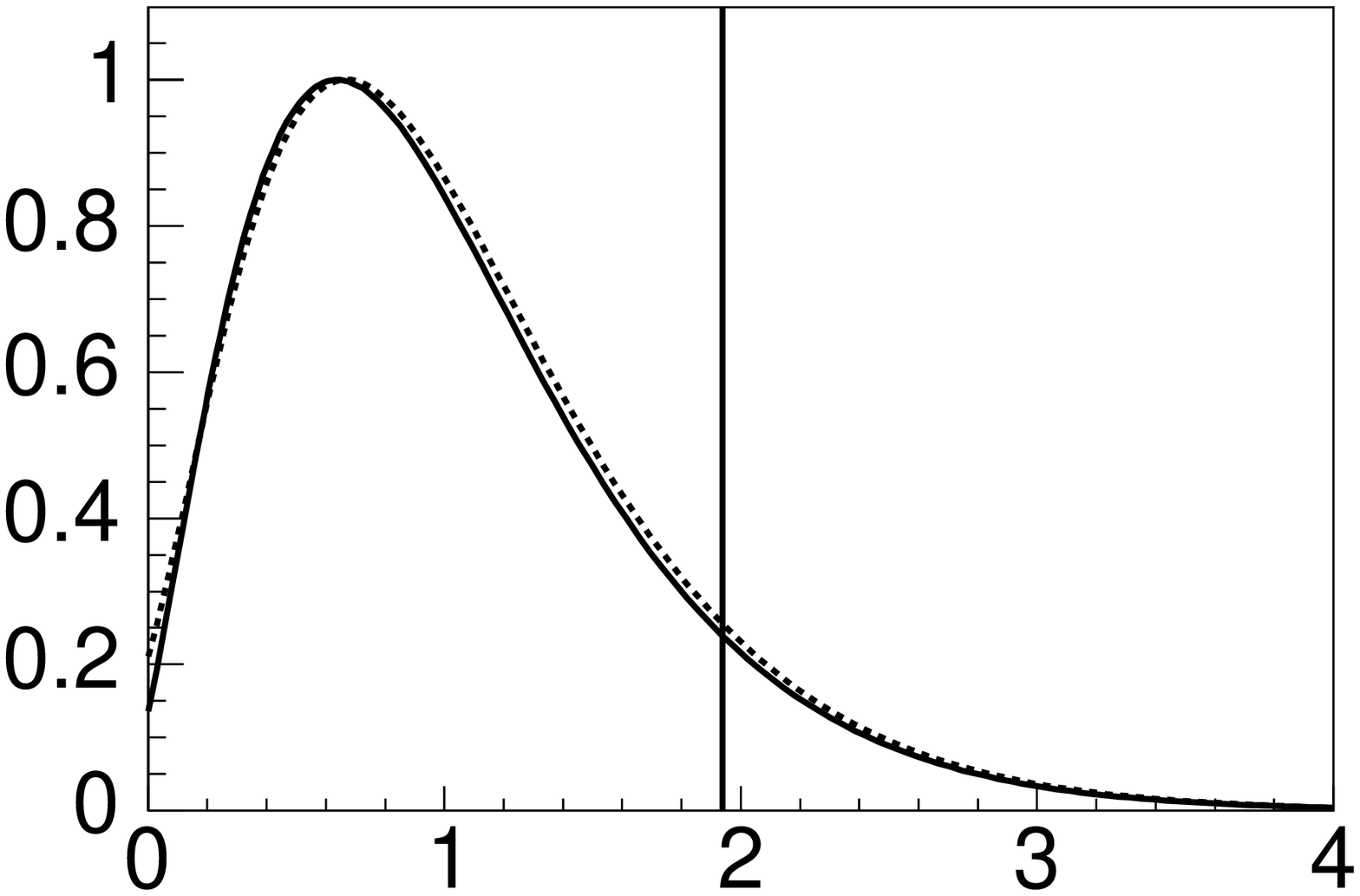,bbllx=0,bblly=0,bburx=555,bbury=380,width=0.51\linewidth}%
    \put(-55,60){\boldmath \large $D^+_s a_2^-$}
    \put(-40,-7){\boldmath \bf ${\cal B} \times 10^4$}

\epsfig{figure=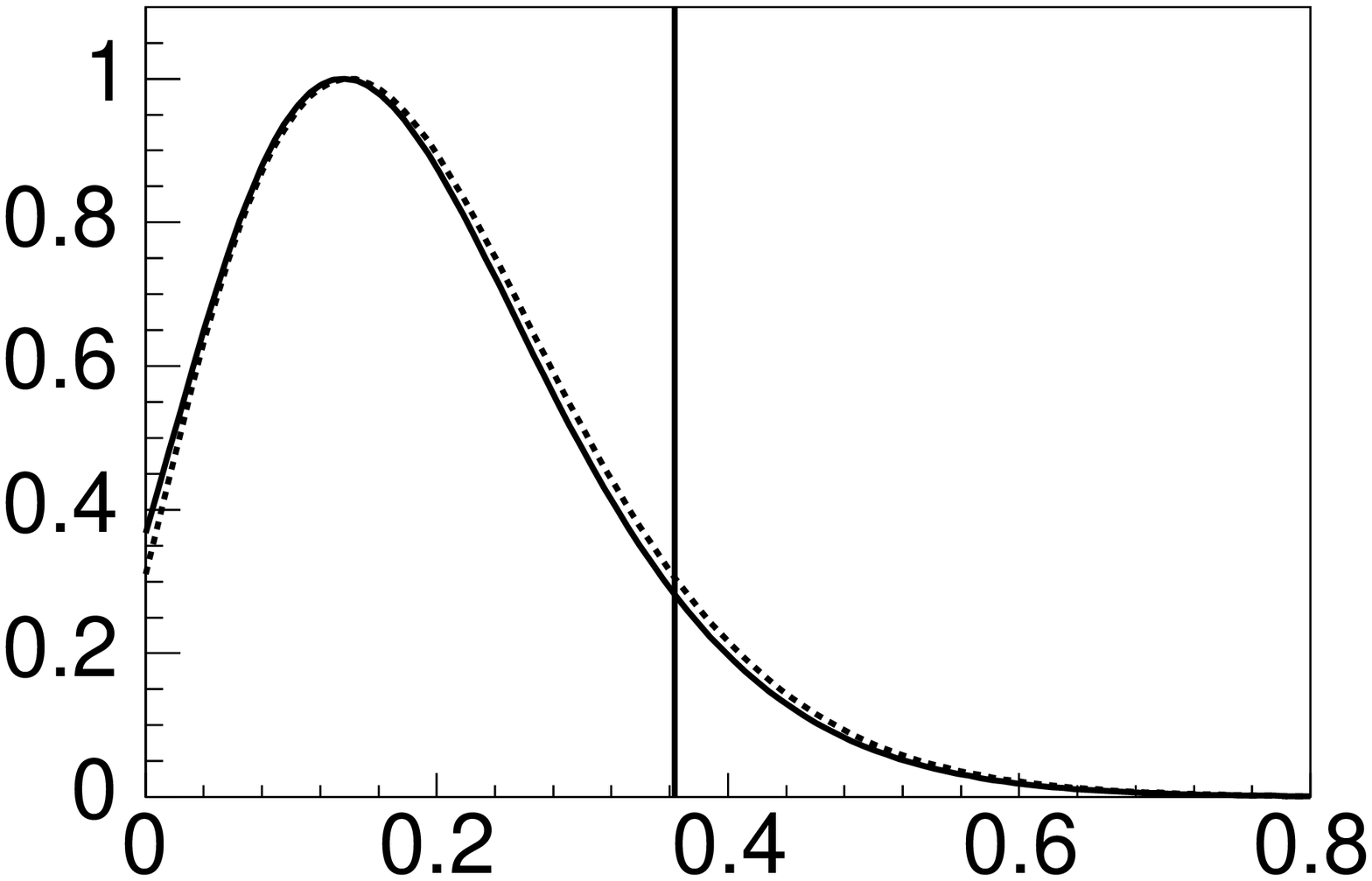,bbllx=0,bblly=0,bburx=555,bbury=385,width=0.51\linewidth}%
    \put(-55,60){\boldmath \large $D^{*+}_s a_0^-$}
    \put(-40,-7){\boldmath \bf ${\cal B} \times 10^4$}
\epsfig{figure=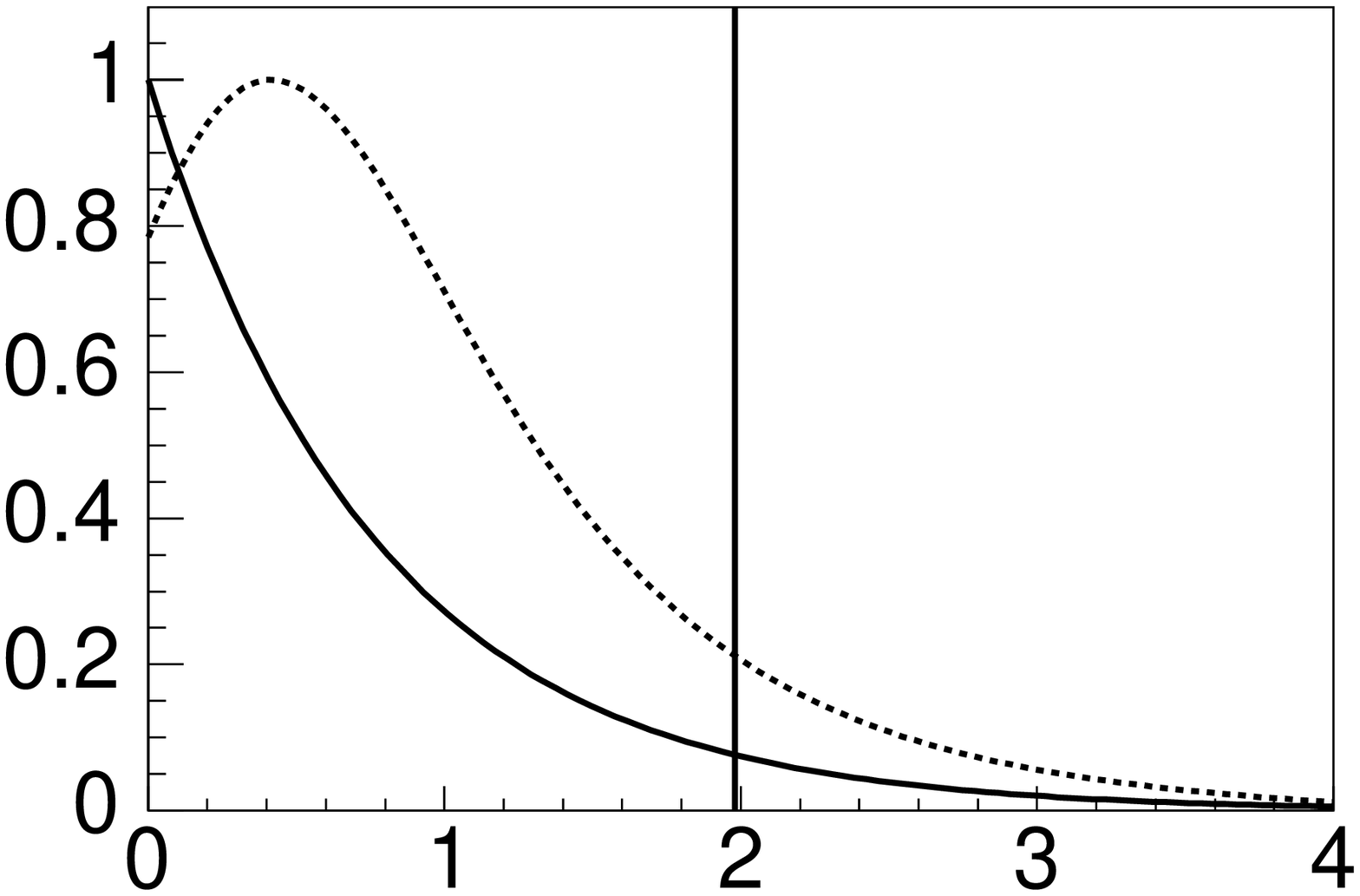,bbllx=0,bblly=0,bburx=555,bbury=385,width=0.51\linewidth}%
    \put(-55,60){\boldmath\large $D^{*+}_s a_2^-$}
    \put(-40,-7){\boldmath \bf ${\cal B} \times 10^4$}
    \caption{
      Likelihood functions of the fit for the \mes\ distributions of the
      selected $B^0\to D_s^{(*)+} a_{0(2)}^-$ candidates. 
      Solid curves represent the original likelihood scan from the fit,
      the dashed lines show the result of the convolution with the systematic
      errors Gaussian. Vertical lines indicate the 90$\%$~Bayesian C.L. upper limit value.}    
    \label{fig:scan-all}
  \end{center}
\end{figure}

We use a Bayesian approach with a flat prior above zero to set 90$\%$ confidence level
upper limits on the branching fractions. In a given mode, the upper limit
on the branching fraction (${\cal B}_{UL}$) is defined by:
\begin{equation*}
\int_0^{{\cal B}_{UL}} {\cal L}({\cal B}) d{\cal B} = 0.9 \times \int_0^{\infty} {\cal L}({\cal B}) d{\cal B}
\label{eq:uplimit}
\end{equation*}
where ${\cal L}({\cal B})$ is the likelihood as a function of the branching fraction 
${\cal B}$ as determined from the \mes fit described above. 
We account for systematic uncertainties 
by numerically convolving ${\cal L}({\cal B})$ with a Gaussian distribution with a width
determined by the relative systematic uncertainty multiplied by the
branching fraction obtained from the \mes fit. 
In cases with asymmetric errors we took the larger for the width of this 
Gaussian function.
In case of $D^{*+}_s a_2^-$ (where no central value
was determined from the fit) we conservatively estimate the absolute systematic
error by taking the numerically calculated $90\%$ confidence level upper limit
(without the systematic uncertainties) instead of the fitted branching fraction.
The resulting upper limits are summarized in Table~\ref{tab:fit} (fourth column).
The likelihood curves are shown in Figure~\ref{fig:scan-all}.

We have also calculated upper limits without including the intermediate
branching fractions of the decays
$D^+_s \ra \phi \pi^+$~\cite{dsphipi-babar} 
and $a_{0(2)}^+ \ra \eta \pi^+$~\cite{PDG2004}.
The relative systematic errors in this case are reduced to $18\%$ for each
of the $B^0$ meson decay modes. The results are presented in Table~\ref{tab:fit}
(third and fourth columns, numbers in parenthesis).

In conclusion, we do not observe any evidence for the decays
$B^0\to D_s^{+} a_0^-$, $B^0\to D_s^{+} a_2^-$, $B^0\to D_s^{*+} a_0^-$ 
and  $B^0\to D_s^{*+} a_2^-$, and set 90\% C.L.\
upper limits on their branching fractions. The upper limit value for
$B^0\to D_s^{+} a_0^-$ is lower than the theoretical expectation,
which might indicate the need to revisit the $B \ra a_0 X$ transition
form factor estimate. It might also imply the limited applicability
of the factorization approach for this decay mode. The upper limits
suggest that the branching ratios of $B^0 \ra D^{(*)+} a_{0(2)}^-$
are too small for $CP$-asymmetry measurements given the present
statistics of the $B$-factories.

We are grateful for the excellent luminosity and machine conditions
provided by our \pep2\ colleagues, 
and for the substantial dedicated effort from
the computing organizations that support \babar.
The collaborating institutions wish to thank 
SLAC for its support and kind hospitality. 
This work is supported by
DOE
and NSF (USA),
NSERC (Canada),
IHEP (China),
CEA and
CNRS-IN2P3
(France),
BMBF and DFG
(Germany),
INFN (Italy),
FOM (The Netherlands),
NFR (Norway),
MIST (Russia), and
PPARC (United Kingdom). 
Individuals have received support from CONACyT (Mexico), A.~P.~Sloan Foundation, 
Research Corporation,
and Alexander von Humboldt Foundation.


\begin{thebibliography}{99}
\bibitem{chargeconj}
Charge conjugate reactions are implicitly included, throughout this paper.
\end{thebibliography}
\end{document}